# Phosphorene: A New 2D Material with High Carrier Mobility

Han Liu[1,2], Adam T. Neal[1,2], Zhen Zhu[3], David Tománek[3], and Peide D. Ye[1,2] *

[1] School of Electrical and Computer Engineering, Purdue University, West Lafayette, Indiana 47907, USA

[2] Birck Nanotechnology Center, Purdue University, West Lafayette, Indiana 47907, USA

[3] Physics and Astronomy Department, Michigan State University, East Lansing, Michigan 48824, USA

* Email of the corresponding author: yep@purdue.edu

**Preceding the current interest in layered materials for electronic applications, research in the 1960's found that black phosphorus combines high carrier mobility with a fundamental band gap.[1] We introduce its counterpart, dubbed few-layer phosphorene, as a new 2D *p*-type material. Same as graphene[2,3] and MoS$_2$,[4] phosphorene is flexible and can be mechanically exfoliated. We find phosphorene to be stable and, unlike graphene, to have an inherent, direct and appreciable band-gap that depends on the number of layers. Our transport studies indicate a carrier mobility that reflects its structural anisotropy and is superior to MoS$_2$. At room temperature, our phosphorene field-effect transistors with 1.0 μm channel length display a high on-current of 194 mA/mm, a high hole field-effect mobility of 286 cm$^2$/V·s, and an on/off ratio up to 10$^4$. We demonstrate the possibility of phosphorene integration by constructing the first 2D CMOS inverter of phosphorene PMOS and MoS$_2$ NMOS transistors.**

Our findings are in-line with the current interest in layered solids cleaved to 2D crystals, represented by graphene and transition metal dichalcogenides (TMDs) such as MoS$_2$, which exhibit superior mechanical, electrical and optical properties over their bulk counterparts and open the way to new device concepts in the post-silicon era.[2-5] An important advantage of these atomically thin 2D semiconductors is their superior resistance to short channel effects at the scaling limit.[6] Massless Dirac fermions endow graphene with superior carrier mobility, but its semi-metallic nature limits seriously its device applications.[7,8] Semiconducting TMDs, such as MoS$_2$, do not suffer from a vanishing gap[9,10] and have been applied successfully in flexible *n*-type transistors[4] that pave the way toward ultimately scaled low-power electronics. Recent studies on MoS$_2$ transistors have revealed good device performance with a high drain current of up to several hundred mA/mm, a sub-threshold swing down to 74 mV/dec, and an $I_{on}/I_{off}$ ratio of over $10^8$.[4,11-13] Due to the presence of S vacancies in the film and the partially Fermi level pinning near the conduction band,[12,14,15] $MoS_2$ transistors show *n*-type FET characteristics. In previously demonstrated $MoS_2$ logic circuits based on *n*-type transistors only, the static power consumption is likely too large for low-power integrated systems.[16,17] This fact alone calls for new *p*-type semiconductors that would allow the realization of CMOS logic in a 2D device. In this study, we introduce phosphorene, a name we coined for a monolayer of black phosphorus, and few-layer phosphorene as novel 2D *p*-type high-mobility semiconductors for CMOS applications. We study the electronic properties, transport behavior, and furthermore demonstrate the first CMOS inverter using few-layer phosphorene as the *p*-channel and $MoS_2$ as the *n*-channel.

Black phosphorus, the bulk counterpart of phosphorene, is the most stable phosphorus allotrope at room temperature[18] that was first synthesized from red phosphorus under high pressure and high temperature in 1914.[19] Similar to graphite, its layered structure is held together by weak inter-layer forces with significant van der Waals character.[20-22] Previous studies have shown this material to display a sequence of structural phase transformations, superconductivity at high pressures with $T_c$ above 10 K, and temperature dependent resistivity and magnetoresistivity.[1,22-27] 2D phosphorene is, besides graphene, the only stable elemental 2D material that can be mechanically exfoliated.

We have determined the equilibrium structure and bonding of black phosphorus and few-layer phosphorene using *ab-initio* density functional theory (DFT) calculations based on the PBE and HSE06 functionals[28] as implemented in the SIESTA[29] and VASP[30] codes. As seen in the optimized structure depicted in Figs. 1a-c, phosphorene layers share a honeycomb lattice structure with graphene with the notable difference of non-planarity in the shape of structural ridges. The bulk lattice parameters $a_1 = 3.36$ Å, $a_2 = 4.53$ Å, and $a_3 = 11.17$ Å, which have been optimized by DFT-PBE calculations, are in good agreement with the experiment. The relatively large value of $a_3$ is caused by the nonplanar layer structure and the presence of two AB stacked layers in the bulk unit cell. The orthogonal lattice parameters $a_1 = 3.35$ Å and $a_2 = 4.62$ Å of the monolayer lattice, depicted in Figs. 1b-c, are close to those of the bulk structure, as expected in view of the weak 20 meV/atom interlayer interaction that is comparable to graphite. We note that the ridged layer structure helps to keep orientational order between adjacent phosphorene monolayers and thus maintains the in-plane anisotropy; this is significantly different from graphene with its propensity to form turbostratic graphite.

Our DFT-PBE band structure results in Fig. 1d indicate that a phosphorene monolayer is a semiconductor with a direct band gap of 0.9 eV at Γ. In agreement with previous studies,[31,32] our calculated band gap for the bulk system is significantly smaller, $E_g = 0.04$ eV. We expect these DFT-based band gap values to be underestimated,[31] as confirmed by comparing to the observed bulk band gap value[1,20,22] of 0.3 eV.

Alternative calculations based on the HSE06 hybrid functional[28], shown for these systems in Fig. 1e, seem to significantly overestimate the band gap.[32] Much more interesting than the precise band gap value is its dependence on the number of layers $N$ in a few-layer slab, shown in Fig. 1e. Our finding that $E_g$ varies significantly between 0.9 eV in a monolayer and 0.1 eV in the bulk promises the possibility of tuning the electronic properties of this system. Equally interesting is the sensitive dependence of the gap on in-layer stress along different directions, shown in Fig. 1f. Of particular importance is our finding that even a moderate ≈3% in-plane compression, possibly caused by epitaxial mismatch with a substrate, will change phosphorene from a direct-gap to an indirect-gap semiconductor with a significantly smaller gap.

Our key finding is the transport behavior of few-layer phosphorene along different directions. Few-layer black phosphorus crystals were peeled and transferred onto a 90 nm $SiO_2$ capped Si substrate. The thickness of the phosphorene film was measured by atomic force microscopy and found to be as low as 1.8 nm, corresponding to a phosphorene tri-layer, as seen in Fig. 2a. Metal contacts were symmetrically defined around the crystal with 45º as the angular increment of the orientation, as shown in Fig. 2b. We fabricated 1 μm wide 20/60 nm thick Ti/Au contacts to few-layer phosphorene so that the spacing between all opposite bars was 5 μm. We used the four pairs of diametrically opposite bars as source/drain contacts for a transistor and measured the transistor behavior for each of these devices. The maximum drain current at 30 V back gate bias and 0.5 V drain bias, which we display in Fig. 2c as a function of the orientation of the contact pair, shows clearly an angle-dependent transport behavior. The anisotropic behavior of the maximum drain current is sinusoidal, characterized by the minimum value of ≈85 mA/mm at 45º and 225º, and the maximum value of ≈137 mA/mm at 135º and 315º. In spite of the limited 45º angular resolution, the observed 50% anisotropy between two orthogonal directions is significant. The same periodic trend can be found in the maximum value of the transconductance, which implies a mobility variation in the x-y plane of few-layer phosphorene. This large mobility variation is rarely seen in other conventional semiconductors. It can be attributed to the uniquely ridged structure in the 2D plane of few-layer phosphorene, seen in Figs. 1a-c, suggesting a different transport behavior along or normal to the ridges.

We have also investigated both the field-effect mobility and the Hall mobility of few-layer phosphorene as a function of temperature and present our results in Fig. 2b and Fig. S1 in the Supplementary Material. Due to the above-mentioned dependence of the carrier mobility on the orientation of different few-layer phosphorene samples, we normalized all our mobility results to the room-temperature value and focus on their dependence on temperature. Our four-terminal mobility measurements, presented in Fig. 2d, show an increase by a factor of 5 of the observed field-effect mobility as the temperature drops from 300K to 10K, due to the strong reduction of phonon scattering at low temperatures. In contrast to this finding, our two-terminal measurements, also reproduced in Fig. 2d, show a mobility decrease at lower temperatures. We believe that the two-terminal measurements do not represent the intrinsic mobility behavior, but rather reflect the temperature dependence of the contact resistance caused by the Schottky barriers at the metal/phosphorene interface. We expect the contact resistance to increase and eventually to dominate at lower temperatures, when thermally assisted tunneling is greatly restrained. The apparent discrepancy between these two sets of data confirms the importance of using four-terminal measurements to extract the intrinsic mobility in order to avoid artifacts associated with contact resistances.

Once we have established the transport behavior of few-layer phosphorene, we proceed to fabricate transistors of this novel 2D material in order to examine its performance in actual devices. We employed the same approach to fabricate transistors with a channel length of 1.0 μm as in our previous transport study. We used crystals of few-layer phosphorene with a thickness between 4~6 nm, corresponding to 8~12 layers, to achieve a better device performance, as described in the Supplementary Material. The *I-V* characteristic of a typical phosphorene-based field-effect transistor for back gate voltages ranging from +30 V to -30 V, shown in Fig. 3a, indicates a reduction of the total resistance with decreasing gate voltage, a clear signature of its *p*-type characteristics. Consequently, few-layer phosphorene is a welcome addition to the family of 2D semiconductor materials, since most pristine TMDs are either *n*-type or ambipolar as a consequence of the energy level of S vacancy and charge neutral level coinciding near the conduction band edge of these materials.[12,15] Only in few cases, *p*-type transistors have been fabricated by externally doping 2D systems using gas adsorption, which is not easily practicable for solid-state device applications.[5,33] The observed linear *I-V* relationship at low drain bias is indicative of good contact properties at the metal/phosphorene interface. We also observe good current saturation at high drain bias values, with the highest drain current of 194 mA/mm at 1.0 μm channel length at the back gate voltage $V_{bg}$=-30 V and drain voltage $V_{ds}$=-2 V. In Fig. 3b we present the transfer curves for drain bias values $V_{ds}$=0.01 V and 0.5 V, which indicate a current on/off ratio of ~$10^4$, a very reasonable value for a material with a bulk band-gap of 0.3 eV. We also note that according to Fig.1e, the band gap of few-layer phosphorene is strongly enhanced due to the vertical quantum confinement.

Inspecting the transfer curves in Fig. 3b, we find the maximum trans-conductance to range from $G_m$=45 μS/mm at $V_{ds}$=0.01 V to 2.28 mS/mm at 0.5 V drain bias. Using simple Square Law Theory, we can estimate the field-effect mobility $\mu_{FE}$ from $G_m = \mu_{FE} C_{ox} \frac{W}{L} V_{ds}$, where $C_{ox}$ is the capacitance of the gate oxide, *W* and *L* are the channel width and length, and $V_{ds}$ is the drain bias. Our results for $V_{ds}$=0.01 V indicate a high field-effect mobility $\mu_{FE}$=286 m$^2$/V·s at room temperature, and our previous four-terminal measurements suggest a factor of 5 improvement at low-temperatures. These values are still smaller than those in bulk black phosphorus, where the electron and hole mobility is ≈1,000 cm$^2$/V·s at room temperature and could exceed 15,000 $cm^2/V\cdot s$ for electrons and 50,000 $cm^2/V\cdot s$ for holes at low temperatures.[34] We consider the following factors to cause the mobility reduction in few-layer phosphorene. *(i)* The exposed surface of few-layer phosphorene is chemically reactive. Chemisorbed species from the process and the environment change the electronic structure and scatter carriers, thus degrading the mobility. *(ii)* In a particular transistor, the current flow may not match the direction, where the material has the highest in-plane mobility. *(iii)* The Schottky barrier at the metal/phosphorene interface induces a large contact resistance within the undoped source/drain regions. We expect that the real mobility of few-layer phosphorene should increase significantly upon appropriate surface passivation and in a high-k dielectric environment.[35] In any case, the fundamental properties of monolayer and few-layer phosphorene bear promise for achieving much higher carrier mobilities than other TMDs.[36]

To further investigate the nature of the metal/phosphorene junction, we used a three-terminal method, similar to the Kelvin probe, to measure the forward-bias *I-V* characteristics of the Ti/phosphorene metal/semiconductor junction[37] at the constant back gate voltage $V_{bg}$=-30 V and display our results in Fig. 3c. These data show an exponential increase in the current $I_f$ as the voltage $V_f$ across the junction increases from 70 mV to 130 mV. In view of the degenerate doping of the phosphorene sample and the exponential *I-V* characteristics across this junction at temperatures as low as 20 K, we conclude that thermal-assisted tunneling through the Schottky barrier is responsible for the transport through the junction. To determine the Schottky barrier height of the Ti/phosphorene contact, we fit the exponential *I-V* characteristics by the equation $I_f = I_s \exp(V_f/\Phi_0)$, where $I_s$ is the characteristic current and $\Phi_0$ the characteristic energy, which characterize transport across the junction at a particular temperature. Fits of the semi-logarithmic plots in a wide temperature range are shown in Fig. 3c. The temperature dependent characteristic current $I_s$ can be furthermore viewed as proportional to $\exp(\Phi_b/\Phi_0)$, where $\Phi_b$ is the height of the Schottky barrier at the metal-semiconductor junction and $\Phi_0$ is a temperature-dependent quantity. This provides a way to use our temperature dependent *I-V* measurements to determine $\Phi_b$ from the slope of the quantity *log* $I_s$ as a function of $1/\Phi_0$. Figure 3d shows the corresponding plot, where each data point has been determined by fitting the *I-V* characteristic curve at a particular gate voltage and temperature. The slope of all curves shows an impressive independence of the measurement conditions, indicating the Schottky barrier height $\Phi_b \approx 0.21$ eV for holes at the Ti/phosphorene junction. We note that the barrier height determined here is the true Schottky barrier height at the metal/phosphorene junction, not an effective Schottky barrier height that is commonly determined for metal/semiconductor junctions via the activation energy method.[12]

Finally, we demonstrate a CMOS logic circuit containing 2D crystals of pure few-layer phosphorene as one of the channel materials. Since phosphorene shows well-behaved *p*-type transistor characteristics, it can complement well *n*-type $MoS_2$ transistors. Here we demonstrate the simplest CMOS circuit element, an inverter, by using $MoS_2$ for the *n*-type transistor and phosphorene for the *p*-type transistor, both integrated on the same $Si/SiO_2$ substrate. Few-layer $MoS_2$ and phosphorene flakes were transferred onto the same substrate successively by the scotch tape technique. Source/drain regions were defined by e-beam lithography, similar to the PMOS fabrication described above. We chose different channel lengths of 0.5 μm for $MoS_2$ and 1 μm for phosphorene transistors, to compensate for the mobility difference between $MoS_2$ and phosphorene by modifying the width/length ratio for NMOS and PMOS. Ti/Au of 20/60 nm were used for both $MoS_2$ and phosphorene contacts. Prior to top growth of a high-k dielectric, a 1 nm Al layer was deposited on the sample by e-beam evaporation. The Al layer was oxidized in ambient conditions to serve as the seeding layer. A 20 nm $Al_2O_3$ grown by atomic layer deposition (ALD) was used as the top gate dielectric. Finally, 20/60 nm Ti/Au was used for the top gate metal electrode and interconnects between the transistors. The final device structure is shown in Fig. 4a and the corresponding circuit configuration in Fig. 4b. In our CMOS inverter, the power supply at voltage $V_{DD}$ is connected to the drain electrode of the phosphorene PMOS. The PMOS source and the NMOS drain are connected and provide the output voltage signal $V_{OUT}$. The NMOS source is connected to the ground (GND). Both top gates of the NMOS and the PMOS are connected to the source of the input voltage $V_{IN}$. The voltage transfer characteristics (VTC) are shown in Fig. 4c. The power supply voltage ranges from 0.5 V to 2.0 V with a 0.5 V step. Within the input voltage range from -5 V to +5 V, the output voltage shows a clear transition from $V_{DD}$ to 0.

In summary, we have investigated the electrical properties and potential device applications of exfoliated few-layer phosphorene films as a new *p*-type semiconducting 2D material with high carrier mobility. We used *ab initio* calculations to determine the equilibrium structure and the inter-layer interaction of bulk black phosphorus as well as few-layer phosphorene with 1-8 layers. Our band structure calculations indicate that few-layer phosphorene is a direct-gap semiconductor with a strong band gap dependence on the number of layers. We also find a direct-to-indirect band gap transition when phosphorene is subject to small compressive stress. We find substantial anisotropy in the transport behavior of this 2D material, which we associate with the unique ridge structure of the layers. We observe an increase in the carrier mobility by a factor of 5 by cooling phosphorene down from room temperature to 10 K. The overall device behavior can be explained by considering a Schottky barrier height of 0.21 eV for hole tunneling at the junctions between phosphorene and Ti metal contacts. We report fabrication of *p*-type transistors of few-layer phosphorene with a high on-current of 194 mA/mm at 1.0 μm channel length, a current on/off ratio over $10^4$, and a high field-effect mobility up to 286 cm$^2$/V·s at room temperature. We have also constructed a CMOS inverter by combining a phosphorene PMOS transistor with a MoS$_2$ NMOS transistor, thus achieving heterogeneous integration of semiconducting phosphorene crystals as a novel channel material for future electronic applications.

**Acknowledgements**

This material is based upon work partly supported by NSF under Grant CMMI-1120577 and SRC under Tasks 2362 and 2396. Theoretical work has been funded by the National Science Foundation Cooperative Agreement #EEC-0832785, titled "NSEC: Center for High-rate Nanomanufacturing". Computational resources have been provided by the Michigan State University High-Performance Computing Center. The authors would like to thank Yanqing Wu, James C.M. Hwang, and Xianfan Xu for valuable discussions.

**Additional information**

The authors declare no competing financial interests. Correspondence and requests for materials should be addressed to Peide D. Ye at yep@purdue.edu.

**Figure Captions**

**Figure 1 | Crystal structure and band structure of few-layer phosphorene. a,** A perspective side view of few-layer phosphorene. **b,c,** Side and top views of few-layer phosphorene. **d,** DFT-PBE band structure of a phosphorene monolayer. **e,f,** DFT-PBE results for the dependence of the energy gap in few-layer phosphorene on (e) the number of layers and (f) the applied strain along the x- and y-direction within a monolayer. Besides DFT-PBE results, we present HSE06 and available experimental results in (e).

**Figure 2 | Transport properties of phosphorene. a,** Atomic Force Microscopy image of an ≈1.8 nm thick crystal containing a phosphorene tri-layer. **b,** Device structure used to determine the angle-dependent transport behavior. **c,** Angular dependence of the drain current and the trans-conductance $G_m$. **d,** Temperature dependence of the field-effect mobility $\mu_{FE}$ obtained using 4- and 2-terminal measurements.

**Figure 3 | Device performance of *p*-type transistors based on few-layer phosphorene. a,b,** Output (a) and transfer (b) curves of a typical phosphorene transistor. **c,** Forward-bias $I_f$-$V_f$ characteristics of the Ti/black phosphorus junction. **d,** Logarithmic plot of the characteristic current $I_s$ as a function of the reciprocal characteristic energy $\Phi_0$, based on data from (c), which is used to determine the Schottky barrier height $\Phi_b$.

**Figure 4 | CMOS logic with 2D crystals. a,** Schematic view of the CMOS inverter,

with MoS$_2$ serving as the NMOS and few-layer phosphorene serving as the PMOS. **b,** Circuit configuration of the CMOS inverter. **c,** Voltage Transfer Curve $V_{out}(V_{in})$ of the 2D CMOS inverter.

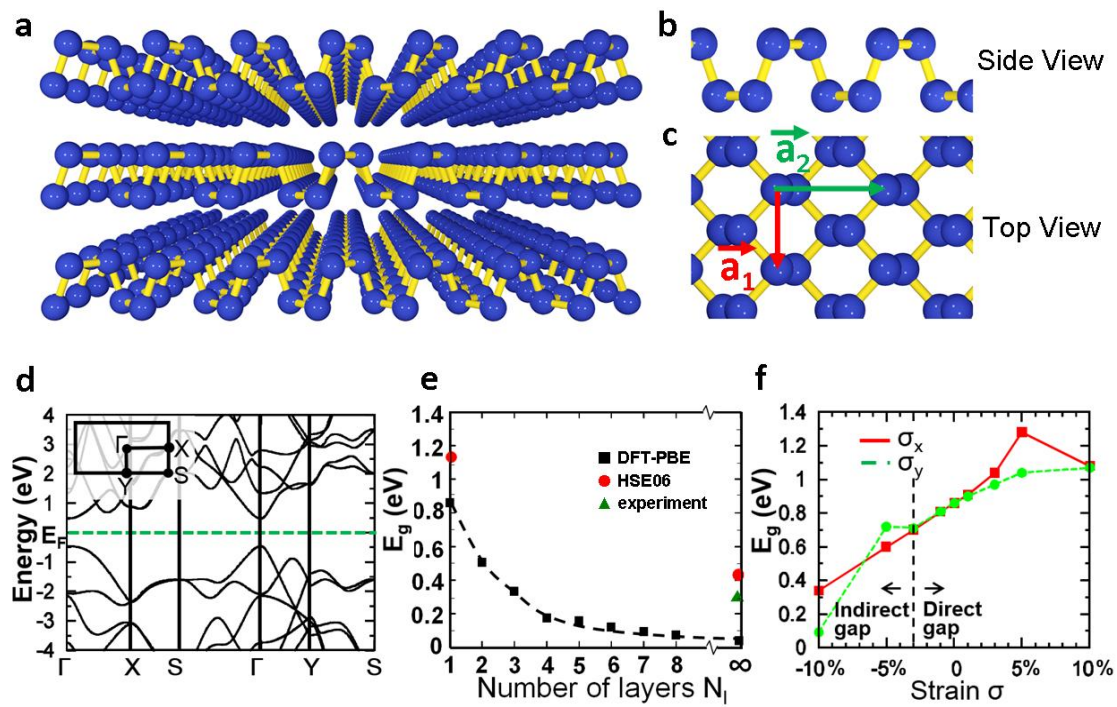

Figure 1

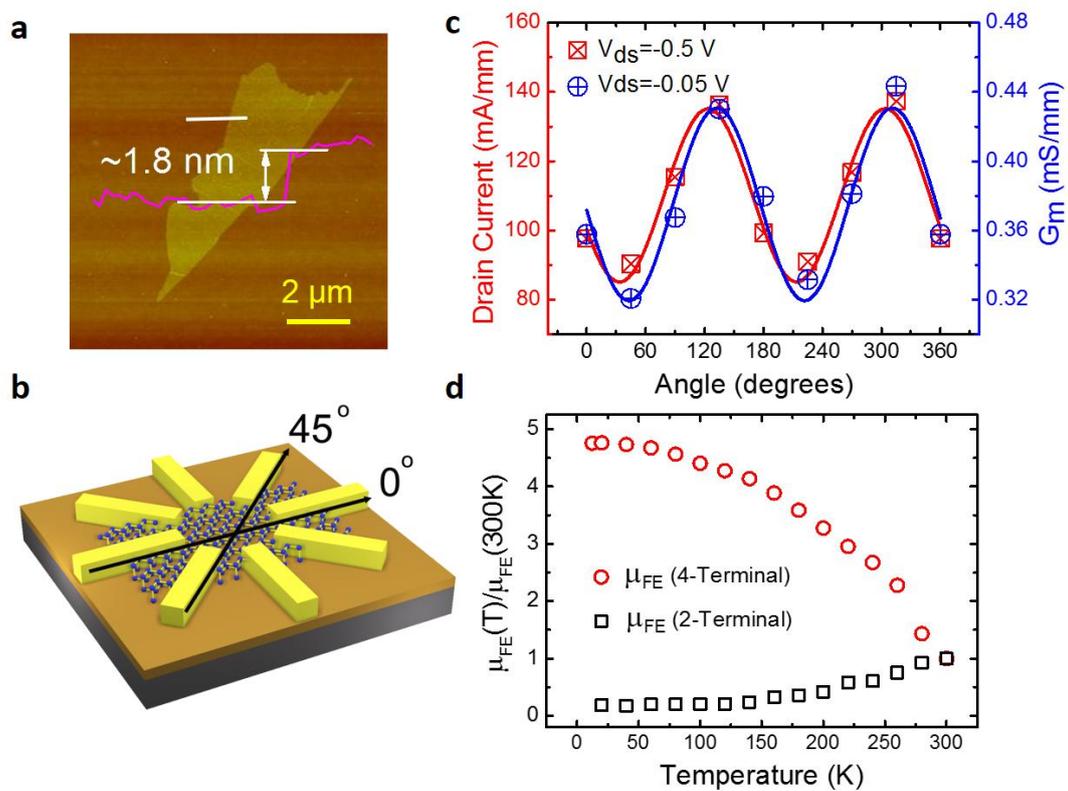

Figure 2

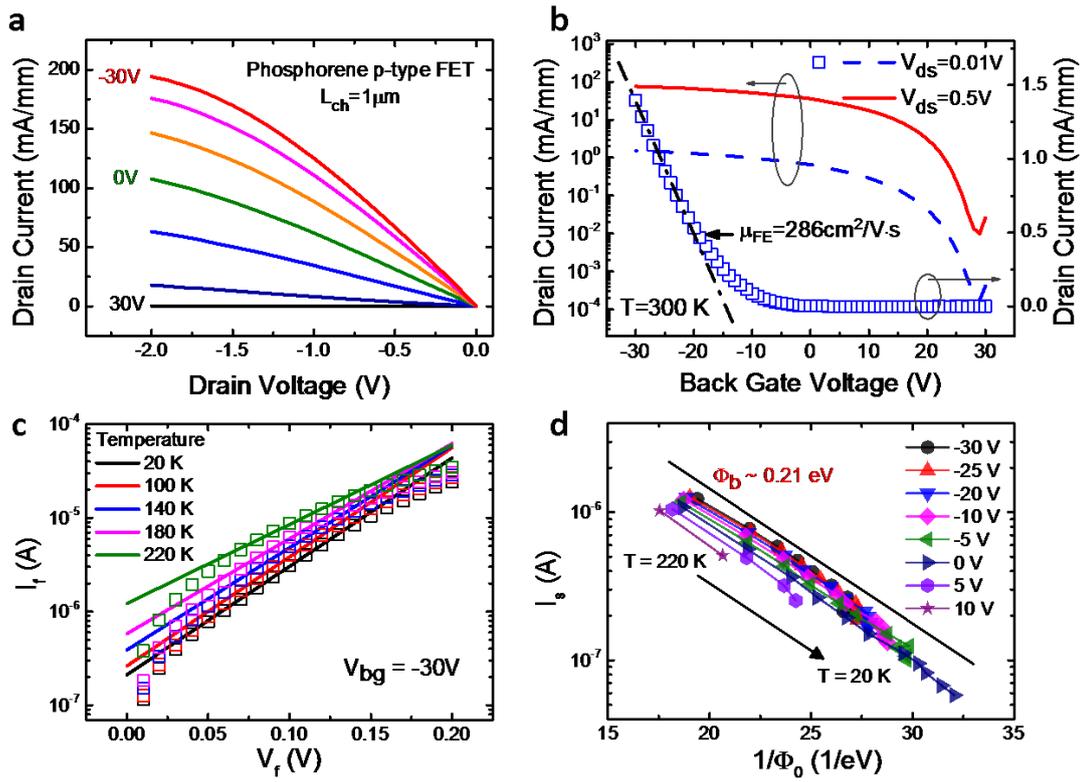

Figure 3

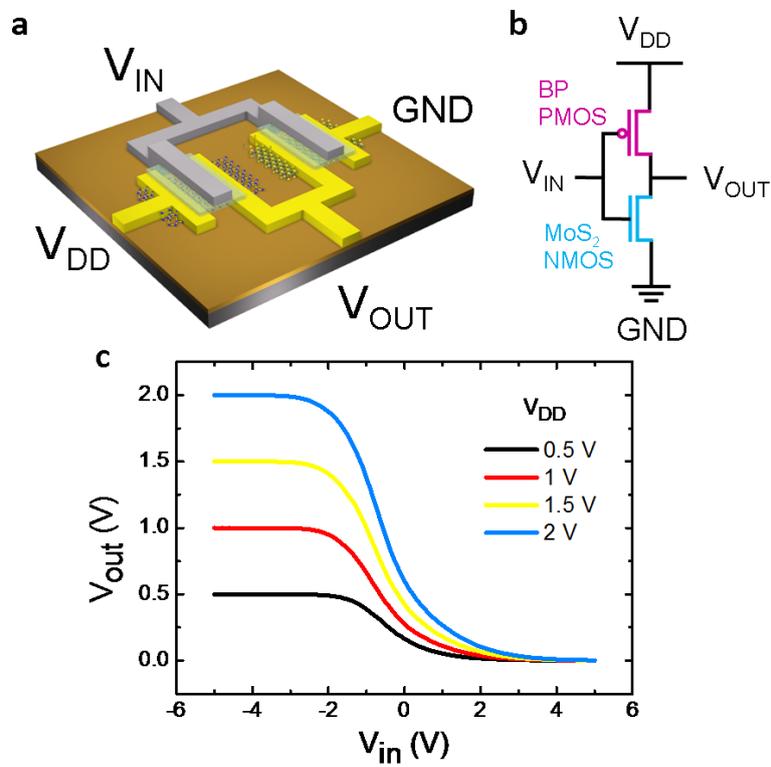

Figure 4

Supplementary Material for

# Phosphorene: A New 2D Material with High Carrier Mobility

Han Liu[1,2], Adam T. Neal[1,2], Zhen Zhu[3], David Tománek[3], and Peide D. Ye[1,2] *

[1] School of Electrical and Computer Engineering, Purdue University, West Lafayette, Indiana 47907, USA

[2] Birck Nanotechnology Center, Purdue University, West Lafayette, Indiana 47907, USA

[3] Physics and Astronomy Department, Michigan State University, East Lansing, Michigan 48824, USA

* Email of the corresponding author: yep@purdue.edu

*Ab initio* **Calculations**

We determined the equilibrium structure, stability and electronic properties of bulk black phosphorus and few-layer phosphorene using *ab initio* density functional theory (DFT) calculations as implemented in the SIESTA code.[1] Few-layer phosphorene crystals have been represented by a periodic array of slabs, separated by a 15 Å thick vacuum region. We used the Perdew-Burke-Ernzerhof (PBE)[2] exchange-correlation functional, norm-conserving Troullier-Martins pseudopotentials[3], and a double-ζ basis including polarization orbitals. The reciprocal space was sampled by a fine *k*-point grid[4] in the Brillouin zone of the primitive unit cell, using the 8×8×2 grid for the bulk and the 8×8×1 grid for few-layer phosphorene. We used a mesh cutoff energy of 180 Ry to determine the self-consistent charge density, which provided us with a precision in total energy of ≤2 meV/atom.

Electronic band structure results obtained by DFT must be interpreted very carefully, since DFT generally underestimates the fundamental band gap. Our calculated bulk band gap value $E_g = 0.04$ eV is in line with previous theoretical results,[5,6] but smaller than the observed bulk band gap value of 0.3 eV. To obtain an independent estimate of the fundamental band gap, we also used the Heyd-Scuseria-Ernzerhof (HSE06)[7] hybrid exchange-correlation functional as implemented in the VASP code.[8] The fundamental band gap value in HSE06 calculations depends sensitively on the mixing between the Hartree-Fock exchange and the PBE exchange-correlation functional. Using the default mixing parameter AEXX=0.25, we obtained $E_g = 0.82$ eV for the fundamental band gap of the bulk material, which is significantly larger than the observed value. Our HSE06 results for the bulk system and monolayer, presented in Fig. 1e of the main manuscript, have been obtained using a reduced value AEXX=0.1 and provide a bulk band gap value of 0.44 eV, which is in closer agreement with the observed value. Whereas HSE06 increases the fundamental band gap value by a similar amount in the monolayer and in the bulk, it also predicts a stress-induced direct-to-indirect gap transition in the monolayer, similar to the DFT-PBE results shown in Fig. 1f of the main manuscript.

**Device Fabrication**

We have started the fabrication of all 2D devices containing phosphorene or $MoS_2$ layers using the scotch-tape based micro-cleavage of the layered bulk black phosphorus and $MoS_2$ crystals, followed by transfer onto the $Si/SiO_2$ substrate, as previously described in graphene studies. Bulk crystals were purchased from Smart-elements (black phosphorus) and SPI Supplies ($MoS_2$). Degenerately doped silicon wafers (0.01-0.02 Ω·cm) capped with 90 nm $SiO_2$ were purchased from SQI (Silicon Quest International). After few-layer crystals of phosphorene and/or $MoS_2$ were transferred onto the substrate, all samples were sequentially cleaned by acetone, methanol and isopropanol to remove any scotch tape residue. This procedure has been followed by a 180ºC post-bake process to remove solvent residue. The thickness of the crystals was determined by a Veeco Dimension 3100 Atomic Force Microscope. E-beam lithography has been carried out using a Vistec VB6 instrument. 20/60 nm

Ti/Au contacts were deposited using the e-beam evaporator at a rate of 1Å/s to define contact electrodes and metal gates. No annealing has been performed after the deposition of the metal contacts. The top gate dielectric material was deposited by an ASM F-120 ALD system at 200ºC, using trimethylaluminium (TMA) and H$_2$O as precursors. The pulse time was 0.8 s for TMA and 1.2 s for water, and the purge time was 5 s for both.

**Hall Mobility Measurement**

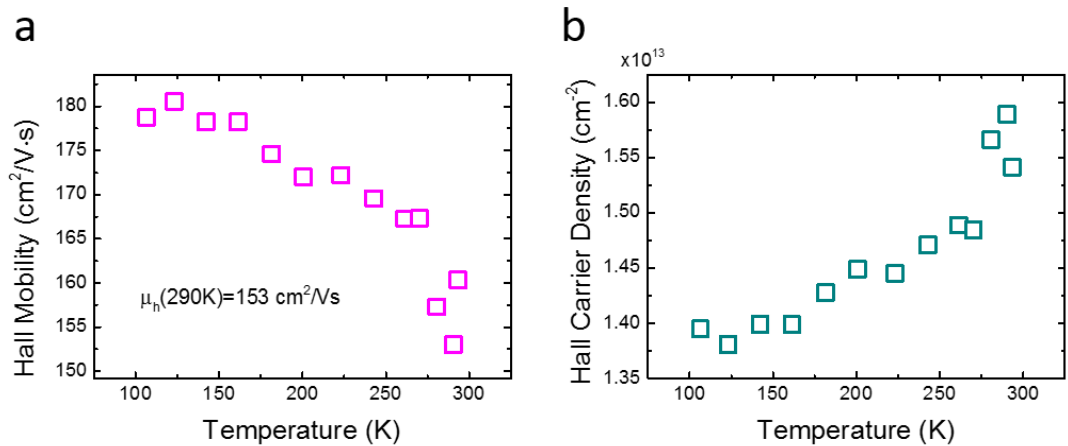

**Figure S1 | Hall mobility measurement. a**, Temperature-dependent Hall mobility and **b**, temperature-dependent Hall carrier density.

The Hall mobility and carrier density of phosphorene between 100K and room temperature have been measured by the standard Hall technique on the fabricated phosphorene Hall-bar structured samples. We have found the room-temperature Hall mobility to be in the same range as the field-effect mobility determined in phosphorene transistors. We also observed an increase in the Hall mobility at lower temperatures, where electron-phonon scattering is suppressed. Since the Hall measurement is, in principle, a 4-terminal measurement, the influence of Schottky contacts should play a less important role in our results.

**Hysteresis in Back-Gated Phosphorene Devices**

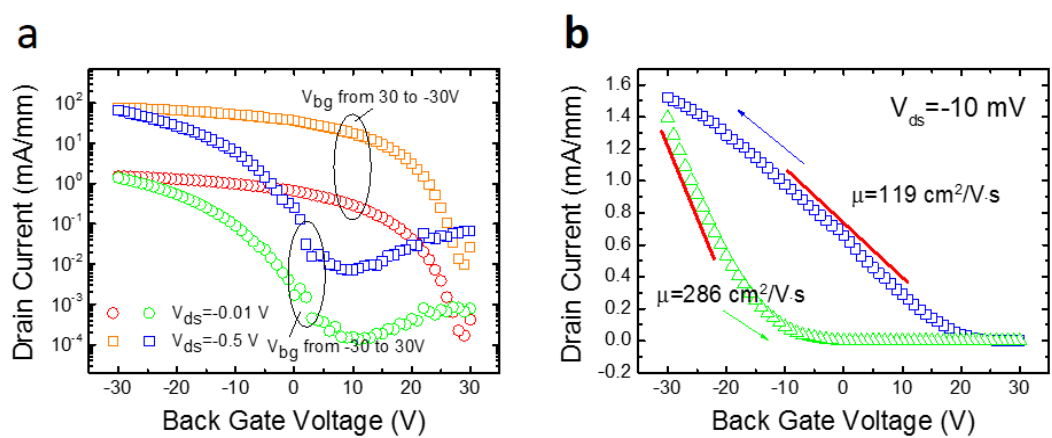

**Figure S2 | Hysteresis in back-gated black phosphorene transistors.** Transfer curves of a back-gated phosphorene transistor presented on a semi-logarithmic scale (a) and a linear scale (b).

In our experiments, back-gated few-layer phosphorene transistors were used to extract the field-effect mobility. The back-gated devices used the globally heavily doped silicon substrate as the gate and 90 nm $SiO_2$ as the gate dielectric. Due to the thick dielectric layer and an un-optimized phosphorene/$SiO_2$ interface, we observed a large hysteresis in bi-directional sweeps of the transfer curves. Though both transfer curves reflect a clear *p*-type transistor performance, a large threshold voltage $V_T$ up-shift of up to 25 V was observed, as seen in Figs. S2a and S2b. The origin of the $V_T$ shift can be mostly attributed to the presence of trapped charges in the dielectric. Because charge traps are charging or discharging at different gate voltages during the sweep from -30 V to +30 V or in the reverse direction, the observed trans-conductance depends on the voltage sweep direction. Therefore, the field-effect mobility, extracted from trans-conductance peak, may show a hysteresis. For the positive sweep, the maximum field-effect mobility is 286 cm$^2$/V·s, whereas it drops to 119 cm$^2$/V·s for the reverse sweep. More serious studies and improvement of the phosphorene passivation and its interface with a high-k dielectric are needed.

**Crystal Thickness of Few-Layer Phosphorene versus Black Phosphorus: Surface Degradation and Current on/off Ratio**

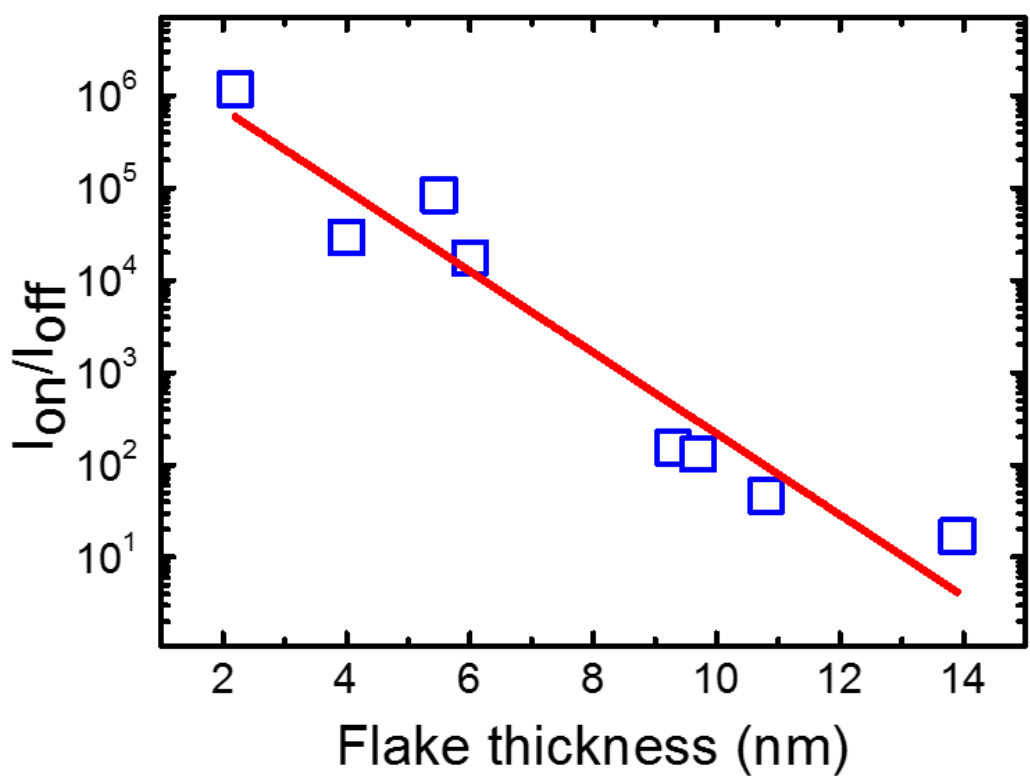

**Figure S3 | On/off current ratio as a function of the few-layer phosphorene crystal thickness.**

We also study the transistor performance as a function of the thickness of few-layer phosphorene crystals and present our results in Fig. S3. We noticed that the crystal thickness underwent a slight reduction during the fabrication process, caused by the chemical reactivity of the impurities, including remaining red phosphorus, in phosphorene. The thickness of the crystals presented in Figure S3 has been measured after the device fabrication. The crystal thickness has a clear impact on the device performance from two aspects. First, it has a strong impact on the current on/off ratio, as shown in Figure S3. We find that devices fabricated on thicker flakes (>8 nm) show a quick drop in the current on/off ratio down to 10-100, whereas devices on thinner crystals have much higher current on/off ratios, which can be up to $10^6$ when the original thickness of the crystal is reduced down to ~2 nm as shown in Figure S3. We associate this behavior mainly with the increased band-gap in few-layer phosphorene films, as seen in Fig. 1e in the main manuscript, due to the vertical quantum confinement. However, we also find a lower on current in thinner crystals containing only few phosphorene layers and, in even thinner crystals, no drain current can be measured. This suggests a trade-off between the device performance and crystal thickness, which is consistent with previous studies on $MoS_2$ transistors.

**Role of Schottky Barriers in Phosphorene Transistors**

Mostly due to lack of source/drain doping, metal contacts on 2D materials usually form considerable Schottky barriers at the metal/semiconductor junction,[9] turning few-layers phosphorene transistors to Schottky transistors instead of conventional

MOSFETs. The presence of Schottky barriers is expected to significantly modify the device parameters and to have a strong impact on the device performance based on the following reasoning.

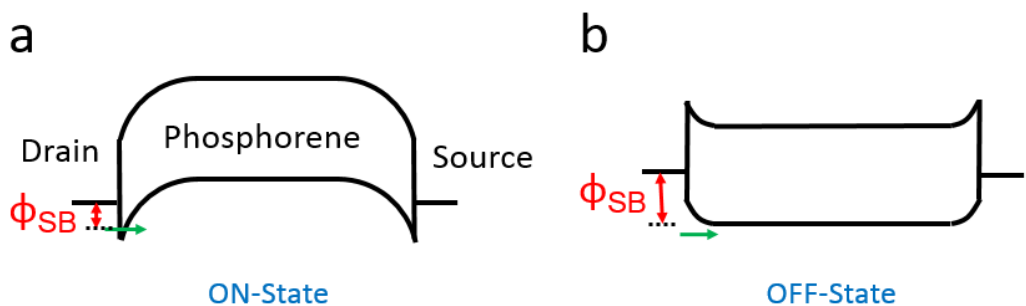

**Figure S4 | Band diagram of phosphorene transistors. a,** In the ON-State, the Schottky barrier width is greatly reduced. **b,** In the OFF-State, the Schottky barrier height is increased.

*(i)* As shown in Figure S4, the ON/OFF states in few-layer phosphorene transistors are not completely controlled by the carrier density in the few-layer phosphorene channel, as is the case in Si MOSFETs. Instead, they are dominantly controlled by the effective Schottky barrier for holes. In the ON-State, e.g. at $V_{bg}$=-30 V, both conduction and valence bands are pushed upwards, leaving a reduced Schottky barrier width, which facilitates the hole injection from the contact metal in a thermally assisted tunneling process. In contrast to this, in the OFF-State, e.g. at $V_{bg}$=+30 V, both bands are pulled down, thus drastically increasing the barrier height for hole tunneling. Therefore, the hole current is strongly suppressed in this case, while the electron current is possibly increased, as also seen in Fig. S2a. Still, the electron current cannot become as large as the hole current, since the mobility of electrons is lower than that of holes and since few-layer phosphorene is a natural *p*-type semiconductor. Therefore, if we still were to use the Drude model to extract the field-effect mobility, the mobility calculated would be underestimated even in a four-terminal measurement (which eliminates the effect of Schottky barriers), since the channel could not be modeled as a pure resistor.

*(ii)* Presence of Schottky barriers at the metal contacts would degrade the performance of transistors and the CMOS inverter as well. In Schottky barrier transistors, it should be more difficult to reach a full saturation of the output curves than in conventional MOSFETs, since a great portion of the drain bias voltage would appear also across the contacts. In conventional Si MOSFET based CMOS inverters, in the ON-state at the vicinity of the switching threshold voltage $V_M$, both the PMOS and the NMOS are in the saturation region, and the drain current is identical in both transistors. A small change in the $V_{in}$ voltage would require the drain current to change drastically. If the transistor is in the saturation region, it requires a drastic change in the drain bias for either transistors, thus to maintain the same drain current for both PMOS and NMOS channels. This drastic change in the drain bias would show up as a large change in $V_{OUT}$, leading to a satisfactory value of the inverter gain. However, in Schottky barrier transistors, which can not easily reach the saturation region, a small change in the drain bias would meet the demand. As a result, the gain of such transistors would be significantly reduced. In our 2D CMOS, most gain values are close to 1. This is caused by the above-mentioned characteristics of contacts at metal/2D semiconductor interfaces. We expect to be able to improve the device performance by reducing the Schottky barrier height/width. We are convinced that contact engineering becomes one of the most important device aspects in 2D materials and devices research.[9]